\documentclass{article}
\usepackage{spconf,amsmath,graphicx,hyperref}

\usepackage{tabularx}
\usepackage{booktabs}

\usepackage{cite}
\usepackage{amssymb,amsfonts}
\usepackage{algorithmic}

\usepackage{textcomp}

\usepackage{xcolor}
\usepackage{subfigure}
\def\BibTeX{{\rm B\kern-.05em{\sc i\kern-.025em b}\kern-.08em
    T\kern-.1667em\lower.7ex\hbox{E}\kern-.125emX}}

\usepackage{spconf,amsmath,graphicx,hyperref}

\title{Improving Audio Event Recognition with Consistency Regularization
}

\twoauthors
  {Shanmuka Sadhu\sthanks{Work done while Shanmuka Sadhu was participating NSF REU Computing for Health and Well-being award \#2349121 at the University of Iowa.}}
    {Dept. of Computer Science\\
     Rutgers University\\
     New Brunswick, NJ, USA\\
     shanmuka.sadhu@rutgers.edu}
  {Weiran Wang}
    {Dept. of Computer Science\\
     University of Iowa\\
     Iowa City, IA, USA\\
     weiran-wang@uiowa.edu}

\begin{document}
\ninept

\maketitle

\begin{abstract}

Consistency regularization (CR), which enforces agreement between model predictions on augmented views,  has found recent benefits in automatic speech recognition~\cite{yao2025crctc}.
In this paper, we propose the use of consistency regularization for audio event recognition,
and demonstrate its effectiveness on AudioSet. 
With extensive ablation studies for both small ($\sim$20k) and large ($\sim$1.8M) supervised training sets, 
we show that CR brings consistent improvement over supervised baselines which already heavily utilize data augmentation, and CR using stronger augmentation and multiple augmentations leads to additional gain for the small training set.
Furthermore, we extend the use of CR into the semi-supervised setup with 20K labeled samples and 1.8M unlabeled samples, and obtain performance improvement over our best model trained on the small set.
Our code is available at \hyperlink{https://github.com/shanmukasadhu/ModifiedAudioMAE}{https://github.com/shanmukasadhu/ModifiedAudioMAE}

\end{abstract}


\section{Introduction}
\label{sec:intro}

Audio Event Recognition (AER) plays a major role in real-world systems, such as wearable devices~\cite{mishra2025spatial}, smart home devices~\cite{vafeiadis2017audio}, surveillance systems~\cite{pages2017homesound,chi2024audio,dacer2025smart}, and human activity~\cite{yuh2021real,seokho2020occupant}.
Many setups, including supervised and self-supervised audio representation learning, have shown significant results on AER datasets. Early success used non pretrained methods\cite{hershey2017cnn}\cite{kong2020large} and CNNs as a base architecture. Later works 
built audio-foundational models such as CAV-MAE\cite{gong2023contrastive}, AudioMAE\cite{huang2022masked}, and CLAP\cite{elizalde2023clap}, often built on vision transformers (ViTs). Although some later pre-trained work was unimodal~\cite{huang2022masked,gong2021ast}, increased success was found with audio-visual learning, such as CAV-MAE \cite{gong2023contrastive} and EquiAV \cite{kim2024equiav}. Additionally, recent advancements in multimodal models such as Large Audio Language Models (LTU~\cite{gong2023listen}, GAMA~\cite{ghosh2024gama,mishra2025spatial}) highlight the importance of audio encoders for various audio understanding tasks. 


CR-CTC~\cite{yao2025crctc} is a recently developed technique for supervised ASR, which enforces consistency between frame-level predictions of augmented views, and improves the alignment/timing of the CTC model. 
In this work, we apply consistency regularization for AER with the AudioMAE architecture~\cite{huang2022masked} on AudioSet, observing consistent improvement. Our contributions include:
\begin{itemize}
    \item We apply CR to supervised 
    AER and demonstrate consistent improvements on both the small 20k training setup and a large 2M training setup over reproduced baselines.
    
    \item Through extensive ablation studies on augmentation methods, we observe that increasing the number of augmentations during training leads to consistent performance gains.
    
    \item We further extend the use of CR to the \emph{semi-supervised} setting, where the 20k set is supervised while the 2M set is unlabeled, applying consistency regularization on both and achieving improved performance over the 20k best model.
\end{itemize}

We achieve an improvement from 37.9 mAP, obtained by a strong baseline~\cite{huang2022masked}, to 39.6 mAP on the 20k setup (a 4.5\% relative improvement); semi-supervised training further increase the accuracy to 40.1 mAP. 
On our 2M setup, we observe an 4.9\% relative improvement from 44.7 to 46.9 mAP.

\section{Related Works}


\textbf{Consistency regularization (CR):} 
Consistency Regularization has been useful in self-supervised methods, such as SimCLR \cite{chen2020a}, BYOL \cite{grill2020bootstrap}, MoCo \cite{he2020momentum} and SimSiam \cite{chen2021exploring}, 
which use different augmented views to learn semantic representations of unlabeled samples. Supervised methods such as RDrop \cite{wu2021rdrop} and cosub \cite{touvron2023cotraining} have 
demonstrated the advantage of CR by encouraging agreement between model predictions on different augmentations. 
In the audio domain, CR-CTC\cite{yao2025crctc} uses frame-level KL-Divergence loss between two augmented views of the same clean input, and demonstrate improvement for supervised ASR. 
For successful application of CR, it is important to study different augmentation techniques to enforce appropriate inductive bias. 

\textbf{Data Augmentation:} 
Data augmentation has been adopted in several modalities 
to improve 
model performance in the supervised setting.
In computer vision, common augmentations include random cropping, random color distortions, and random Gaussian blur\cite{salamon2017deep}.
In this work, we treat input spectrogram as an image and apply vision transformer encoder to it, and find the random erasing technique~\cite{zhong2020random}, originally developed in the vision field, to be sometimes useful.
In the audio domain, 
 vocal tract length perturbation~\cite{jaitly2013vocal}, speed perturbation~\cite{ko2015audio}), and SpecAugment (including Time warping, Time masking, and Frequency Masking) have been proposed.

 

In this work, we use mixup~\cite{zhang2017mixup}, implemented at the spectrogram level, to create more varieties of audio event combinations and therefore improve generalization. 




Masked prediction~\cite{devlin2018bert,he2022masked,hsu2021supervised}, with masking performed either at the input or feature levels, is shown to be a powerful learning paradigm across modalities. In this work, we use the model architecture of AudioMAE~\cite{huang2022masked}, and its pre-trained checkpoint (with masked reconstruction loss on AudioSet) as initialization. Following their implementation, we randomly drop 20\% of the patches when training the ViT architecture. 

\begin{figure*}[t]
    \centering
    \includegraphics[width=\textwidth]{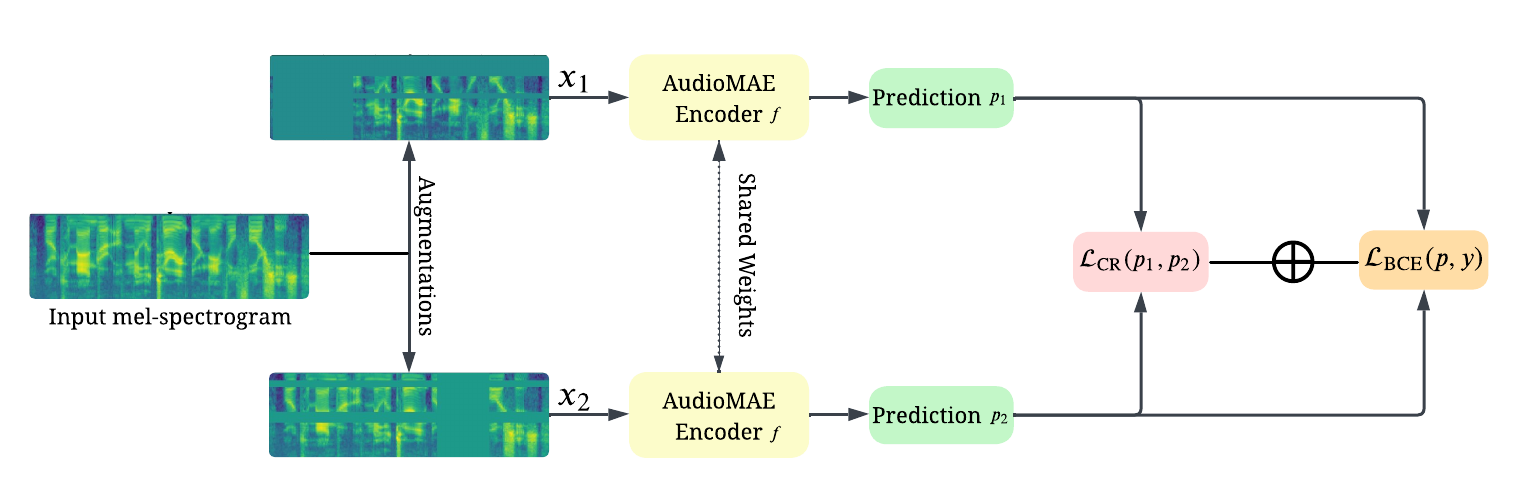}
    \vspace*{-7ex}
    \caption{Overall Architecture of our methodology with 2 augmentations.
     }
    \label{fig:model}
\end{figure*}

\section{Methods}
\subsection{Audio Event Recognition}

Audio Event Recognition (AER) is inherently a multi-label problem, i.e., each audio recording can be associated with multiple classes. We use binary cross entropy (BCE) loss for each class to predict its presence.
For a training set comprising $N$ samples and $C$ events, the BCE loss is defined as
\begin{equation}
\mathcal{L}_{\text{BCE}} = -\frac{1}{N} \sum_{i=1}^{N} \sum_{c=1}^{C} \left[ y_{i,c} \log(\hat{y}_{i,c}) + (1 - y_{i,c}) \log(1 - \hat{y}_{i,c}) \right]
\end{equation}
where $y_{i,c}\in \{0, 1\}$ denotes the target label of class $c$ for sample $i$, and $\hat{y}_{i,c}$ is the predicted probability for class $c$. 
Here logits ($\log \hat{y}_{i,c}$) are obtained by adding a linear layer on top of the global representation of the audio, computed with the ViT encoder followed by average pooling in time.

\subsection{Consistency Regularization}
The intuition behind consistency regularization (CR) is that, by minimizing the KL-Divergence between the model's probability distributions of 2 augmented views, the model learns to focus on invariant/semantic features. 

Suppose $x$ is the input clean spectrogram.
We perform 2 independent augmentations on $x$ to get $x_1$ and $x_2$, 
and apply our model to both $x_1$ and $x_2$ to get two prediction probabilities $p_1$ and $p_2$ respectively. 
The CR loss treats each prediction as ``pseudo-label'' and requires the other prediction to match it, leading to the following cross-entropy loss terms:
\begin{gather}
\mathcal{L}_{\text{p1p2}} = sg(p_1) * \log(p_2) - (1-sg(p_1)) * \log(1-p_2) \\
\mathcal{L}_{\text{p2p1}} = sg(p_2) * \log(p_1) - (1-sg(p_2)) * \log(1-p_1)
\end{gather}
where $sg(\cdot)$ is the stop-gradient operator.
The CR loss function is 
then defined as the average of the above terms:
\begin{equation} \label{eqn:cr-loss}
\mathcal{L}_{\text{CR}} = \frac{1}{2}(\mathcal{L}_{\text{p1p2}}+\mathcal{L}_{\text{p2p1}})
\end{equation}

While the original CR-CTC paper\cite{yao2025crctc} focused on using two augmentations, CR can be extended to multiple views by performing Consistency Regularization pairwise and averaging. Given k augmentations, we can generalize consistency regularization loss as 
\begin{equation}
\mathcal{L}_{\text{CR}} 
=
\frac{1}{k(k-1)} \sum_{i \neq j}
\left( \mathcal{L}_{\text{pipj}}+\mathcal{L}_{\text{pjpi}}
\right)
\end{equation}

For supervised learning of AER, the total loss is 
\begin{equation} \label{eqn:sup-loss}
\mathcal{L}_{\text{total}} =  \mathcal{L}_{\text{BCE}} + \lambda\mathcal{L}_{\text{CR}} 
\end{equation}
where $\lambda > 0$ is a hyper-parameter that controls the consistency regularization strength and is tuned in the ablation study.

We note that the definition of CR loss~\eqref{eqn:cr-loss} does not require the ground truth label. This makes it possible to use it on unsupervised data, as we will demonstrate later.

\subsection{Augmentations}

We now discuss the list of augmentations found to be useful in CR loss,
in the same order they are applied in the data pipeline.

\textbf{Mixup \cite{zhang2017mixup}}:
Mixup is a data-augmentation technique that takes two fbanks features $x_1$ and $x_2$ and creates a new sample $\hat{x}$ that is a convex combination of the two; corresponding labels are mixed using the same coefficient to become the label of $\hat{x}$. Formally, we have
\begin{align*}
\hat{x} & = \mu x_1 +(1-\mu) x_2 \\
\hat{y} & = \mu y_1 +(1-\mu) y_2
\end{align*}
where $\mu > 0$ is randomly sampled from a beta distribution. 
$(\hat{x},  \hat{y})$ is then used as a new sample for supervised learning, increasing the data diversity.

\begin{figure}[t]
    \centering
    \begin{tabular}{@{}c@{\hspace*{0\linewidth}}c@{\hspace*{0\linewidth}}c@{\hspace*{0\linewidth}}c@{\hspace*{0\linewidth}}c@{}}
    \includegraphics[width=0.20\linewidth]{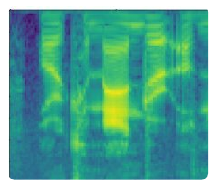}    &  
    \includegraphics[width=0.2\linewidth]{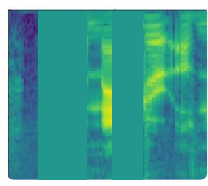}       &
    \includegraphics[width=0.2\linewidth]{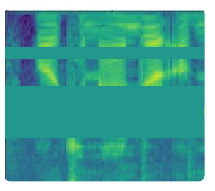}      & 
    \includegraphics[width=0.21\linewidth]{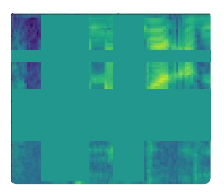}
    & 
    \includegraphics[width=0.2\linewidth]{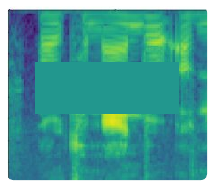}
    \\
    (a) & (b) & (c) & (d) & (e)
    \end{tabular}
    \label{fig:four_images}
    \caption{Illustration of SpecAugment techniques. (a) Original, (b) Time-masked, (c) Frequency-masked, and (d) Time + Frequency Maske, (e) Random Erasing.}
    \label{fig:specaug}
\end{figure}

\textbf{SpecAugment\cite{park2019specaugment}}:
SpecAugment is a common augmentation strategy in the audio domain. As illustrated in Figure~\ref{fig:specaug}, it performs time-masking (b), frequency-masking (c), and time + frequency-masking (d) on log-mel spectrograms. 

\textbf{Random Erasing \cite{zhong2020random}}: Although Random Erasing is a computer vision augmentation technique, we applied it after SpecAugment. 
Random Erasing randomly erases a rectangle region in an image with random values. When perfomring random erasing, scale, aspect ratio, and probability of erasing can be controlled.
We only tune the erasing probability in our experiments. 


\section{Experiments}

\subsection{Experimental Setup}

\textbf{AudioSet: } AudioSet\cite{gemmeke2017audio} is a multi-label audio dataset with ~2 million audio clips of length 10 seconds, totalling to 5,800 hours. 
There are two commonly used supervised training sets:
AS-20k is class-wise balanced and originally contained 22,176 audio samples\cite{he2022masked}, and AS-2M is unbalanced and originally had 2,042,985 audio samples. Due to YouTube restrictions, our AS-20k and AS-2M datasets contain 20,550 and 1,783,977 samples, respectively. 
We use the AS-20k as the small supervised set, and combine the unbalanced and balanced datasets as the large AS-2M training set.
Additionally, our test set contains 18,858 samples (while the original AudioSet test set had 20,383 samples\cite{he2022masked}). Our development set consists of 18,610 samples and is separate from our training sets.

\textbf{Baseline:} We use the same architecture and pretrained model as AudioMAE. We have reproduced their AS-20k result, but 
obtained slightly worse results for the large training set since we had less training data (
$\sim$200K for AS-2M)~\cite{huang2022masked}.  

\textbf{Architecture:}
For the encoder portion of the model (Fig~\ref{fig:model}), we use 12-layer ViT-B Transformer model. There are a total of 88.9M 
trainable parameters. We then use the checkpoint provided by AudioMAE~\cite{huang2022masked}, which is pre-trained on AudioSet 2M, for fine-tuning. 
 Torchaudio is used to compute Kaldi-compatible fbank features as model input. We use a weighted random sampler for training, which samples a balanced sample per epoch as done by AudioMAE. 
 For the AS-20K setup, we use a batch size of 64, Adam optimizer with learning rate of 1e-3, and 60 epochs. For AS-2M setup, we use a batch size of 512, learning rate of 2e-4, and 60 epochs.

\textbf{Label Set:} Evaluation on AudioSet is usually based on 527 classes~\cite{kong2020large}.
However, in view of the label ontology\cite{gemmeke2017audio}, most classes have parent nodes. We add intermediate labels that were located on the path from the root to the original 527 labels. This increases our label count from 527 to 556.
While training is performed on 556 classes, evaluation is still calculated on the commonly used 527 classes to stay consistent with related works. Adding these labels provides the model with more fine-grained label information, and we obtain a small improvement in performance: We achieve 35.8 test mAP when training of 527 classes for AS-20k, and 37.9 mAP when training on the intermediate classes as well (shown in Table~\ref{tab:performance}, Baseline with AudioMAE pretraining).
\textbf{Evaluation Metric: }
We use Mean-Average Precision (mAP) averaged over 527 classes, 
as is commonly done in prior works.




\subsection{Supervised training on AS-20K}
\label{sec:expts-20k}

We now perform an ablation study on the AS-20k to understand the effect of each parameter on our supervised setup. 

\begin{table}[t]
\centering
\caption{Ablation Study for 20k. mAPs are measured on dev set.}
\label{tab:ablation-20k}

\begin{tabular}{@{}cc@{}}
\begin{tabular}{lc}
\toprule
\multicolumn{2}{c}{\textbf{a. CR Coefficient $\lambda$}} \\
\midrule
\textbf{$\lambda$} & \textbf{mAP} \\
\midrule
0 & 34.7 \\
1.5 & 35.7 \\
\textbf{2.0} & \textbf{35.8} \\
2.5 & 35.7 \\
3.0 & 35.5 \\
\bottomrule
\end{tabular}
&
\begin{tabular}{lc}
\toprule
\multicolumn{2}{c}{\textbf{b. Mixup Ratio $\mu$}} \\
\midrule
\textbf{Ratio} & \textbf{mAP} \\
\midrule
0 & 35.2 \\
0.15 & 35.5 \\
0.25 & 35.7 \\
\textbf{0.5} & \textbf{35.8} \\
0.75 & 35.6 \\
\bottomrule
\end{tabular}
\end{tabular}

\vspace{1em} 

\begin{tabular}{@{}cc@{}}
\begin{tabular}{lc}
\toprule
\multicolumn{2}{c}{\textbf{c. Random Erasing}} \\
\midrule
\textbf{Prob.} & \textbf{mAP} \\
\midrule
0 & 35.8 \\
0.15 & 35.9 \\
\textbf{0.25} & \textbf{36.0} \\
0.5 & 35.8 \\

\bottomrule
\end{tabular}
&
\begin{tabular}{lc}
\toprule
\multicolumn{2}{c}{\textbf{d. \# Augmentations}} \\
\midrule
\textbf{\#} & \textbf{mAP} \\
\midrule
2 & 36.0 \\
4 & 35.8 \\
\textbf{6} & \textbf{36.2} \\
7 & 36.0 \\
\bottomrule
\end{tabular}
\end{tabular}
\smallskip
\begin{minipage}{\textwidth}
\vspace{1em}
\end{minipage}
\end{table}

\par\textbf{Consistency Coefficient:} We first perform an ablation on the CR loss coefficient $\lambda$ in~\eqref{eqn:sup-loss}. All experiments in Table~\ref{tab:ablation-20k} (a) perform 2 augmentations.
A consistency coefficient of 2 resulted in the best results: 35.8 mAP. For comparision, we provided the pure baseline of our architecture on AS-20K (corresponding to $\lambda=0$); our best coefficient results in a 1.1 mAP increase. 

\par\textbf{Mixup:}
For this ablation study, we investigate how much mixup will improve mAP performance. Since we observe $\lambda=2$ as the best consistency coefficient, we use it for Table~\ref{tab:ablation-20k} (b). In \cite{huang2022masked}, the mixup ratio of 0.5 was used, 
and we confirm that it is also the optimal for our AS-20k setup, leading to a 0.6 mAP increase compared to not using it. This performance improvement may be due to 
the model being exposed to a greater variety of event combinations.

\par\textbf{Random Erasing:} Although Random Erasing is not common in audio, we investigated its performance in our setup. 
{For our Random Erasing ablation, we keep the aspect ratio fixed at (0.3,0.3) and scale fixed to (0.02, 3.3), while changing the probability of random erasing occurring.}
As shown in Table~\ref{tab:ablation-20k} (c), we observe that a random erasing probability 0.25 gives us a 0.2 mAP improvement, on top of the optimal $\lambda$ and mixup.

\par\textbf{Number of Augmentations:}
For this experiment, we fix the other 3 parameters to be the best performing based off the previous ablation studies. The original consistency regularization performs KL-Divergence on 2 augmentations, but we 
find that 6 augmentations resulted in an additional 0.2 mAP improvement for the 20k training setup, as shown in Table~\ref{tab:ablation-20k} (d).

\subsection{Supervised training on AS-2M}

\begin{table}[t]
\centering
\caption{Ablation Study for AS-2M. mAPs are measured on dev set.}
\label{tab:ablation-2M}

\begin{tabular}{@{}cc@{}}
\begin{tabular}{lc}
\toprule
\multicolumn{2}{c}{\textbf{a. CR coefficient $\lambda$  (no erasing, 2 augmentations)}} \\
\midrule
\textbf{$\lambda$} & \textbf{mAP} \\
\midrule

0 & 44.8 \\
1.0 & 46.3 \\
1.5 & \textbf{46.6} \\
2.0 & 46.5 \\
2.5 & 46.4 \\

\bottomrule
\end{tabular}
\end{tabular}

\vspace{1em} 

\begin{tabular}{@{}cc@{}}
\begin{tabular}{lc}
\toprule
\multicolumn{2}{c}{\textbf{b. Random Erasing}} \\
\midrule
\textbf{Prob.} & \textbf{mAP} \\
\midrule
\textbf{0} & \textbf{46.6} \\
0.25 & 46.5 \\
0.5 & 46.3 \\

\bottomrule
\end{tabular}
&
\begin{tabular}{lc}
\toprule
\multicolumn{2}{c}{\textbf{c. \# of Augmentations}} \\
\midrule
\textbf{\#} & \textbf{mAP} \\
\midrule
\textbf{2} & \textbf{46.6} \\
3 & 46.3 \\
4 & 46.3 \\

\bottomrule
\end{tabular}
\end{tabular}
\smallskip
\begin{minipage}{\textwidth}
\vspace{1em}
\end{minipage}
\end{table}

We perform similar ablation studies for AS-2M as done for AS-20k, except that we keep mixup fixed at 0.5. As seen in Table~\ref{tab:ablation-2M} (a), we obtain a 1.8 mAP improvement with the optimal consistency regularization setup ($\lambda=1.5$). 
Unlike AS-20k, we do not observe any increase in performance when using random erasing. Finally, using more than 2 augmentations is not beneficial for AS-2M. 

\subsection{Semi-supervised learning}

Since consistency regularization can be computed without ground-truth labels, we experiment with a semi-supervised setup where we assume the AS-20k data as supervised while the AS-2M data as unsupervised, and apply consistency regularization on both. 

Our training loss for this setup is
\begin{align}
\mathcal{L}_{\text{semi}} = \mathcal{L}_{\text{BCE(20k)}}+\lambda_1\mathcal{L}_{\text{CR(20k)}}+\lambda_2\mathcal{L}_{\text{CR(2M)}}
\end{align}
where $\lambda_1$ is the coefficient for the AS-20k consistency regularization and $\lambda_2$ is the coefficient for the AS-2M consistency regularization.

For every training step, 4 times more unlabeled data is used than labeled data. Thus, in 60 epochs we will see the 1.8M unsupervised set around 3 times.
We inherit the random erasing and mixup strategy for the 20K supervised portion from Section~\ref{sec:expts-20k}. 
We find the most optimal coefficient setup to be 1.5 for \textcolor{blue}{$\lambda_1$} and 1.0 for 
$\lambda_2$, as shown in Table~\ref{tab:ablation-semi} (a), while random erasing and mixup are not applied to unlabeled data.
As shown in Table~\ref{tab:ablation-semi} (b) and (c), adding these augmentations to unlabeled data degrades performance. 
Overall, we achieve 0.4 mAP gain over the best supervised model on AudioSet 20K (cf Table~\ref{tab:ablation-20k}).


\begin{table}[t]
\centering
\caption{Ablation Study for Semi-Supervised Setup on unlabeled dataset. mAPs are measured on dev set.}
\vspace{1em} 
\label{tab:ablation-semi}

\begin{tabular}{@{}cc@{}}

\begin{tabular}{lcc}
\toprule
\multicolumn{2}{c}{\textbf{a. Loss Coefficients (w. optimal aug.)}} \\
\midrule
\textbf{$\lambda_1$} & \textbf{$\lambda_2$} & \textbf{mAP} \\
\midrule
0 & 2.0 & 35.8 \\
1 & 1 & 36.4 \\
1 & 1.5 & \textbf{36.6} \\
1 & 2.0 & 36.5 \\
1.5 & 1 & 36.3 \\
1.5 & 1.5 & 36.4 \\

\bottomrule
\end{tabular}
\end{tabular}

\vspace{1em} 

\begin{tabular}{@{}cc@{}}
\begin{tabular}{lc}
\toprule
\multicolumn{2}{c}{\textbf{b. Random Erasing (Unsup)}} \\
\midrule
\textbf{Prob.} & \textbf{mAP} \\
\midrule
0 & 36.4 \\
\textbf{0.25} & \textbf{36.6} \\

0.5 & 36.0 \\

\bottomrule
\end{tabular}
&
\begin{tabular}{lc}
\toprule
\multicolumn{2}{c}{\textbf{c. Mixup Ratio $\mu$ (Unsup)}} \\
\midrule
\textbf{Ratio} & \textbf{mAP} \\
\midrule
\textbf{0} & \textbf{36.6} \\
0.25 & 36.4 \\
0.5 & 36.4 \\

\bottomrule
\end{tabular}
\end{tabular}
\smallskip
\begin{minipage}{\textwidth}
\vspace{1em}
\end{minipage}
\vspace*{-2ex}
\end{table}


\subsection{Final Results}

\begin{table}[t]
\centering
\caption{Comparison of AER Models. mAPs measured on test set.
}
\label{tab:performance}
\begin{tabular}{lcc}
\toprule
\textbf{Model} & \textbf{AS-20k (mAP)} & \textbf{AS-2M (mAP)} \\
\midrule
PANNs\cite{kong2020large} & 27.8 & 43.1 \\
AST\cite{gong2021ast} & 37.8 & 48.5 \\
AudioMAE\cite{huang2022masked} & 37.1 & 47.3 \\
SSLAM\cite{alex2025sslam} & 40.9 & 50.2 \\

\midrule
\multicolumn{3}{c}{With AudioMAE pretraining} \\
Baseline & 37.9 & 44.7\textsuperscript{*} \\
Ours, Supervised & \textbf{39.6} & \bf 46.9\textsuperscript{*} \\
Ours, Semi-Supervised & \multicolumn{2}{c}{40.1} \\

\midrule
\multicolumn{3}{c}{Without pretraining} \\
Baseline & 17.2 & 30.9\textsuperscript{*} \\
Ours, Supervised & \textbf{19.3} & \bf 33.5\textsuperscript{*} \\
Ours, Semi-Supervised & \multicolumn{2}{c}{19.9} \\

\bottomrule
\end{tabular}

\smallskip
\begin{minipage}{\textwidth}
\textsuperscript{*}Model  trained with 1.8M samples rather than 2M training set size.
\end{minipage}
\end{table}

Table~\ref{tab:performance} compares our proposed approaches (supervised and semi-supervised) to our baseline 
as well as other prevalent AER models. 

Our baseline is (reproduced) AudioMAE~\cite{huang2022masked} without consistency regularization. As shown in Table~\ref{tab:performance}  (middle section), starting from the pretrained checkpoint, we obtain 1.7 mAP improvement in the 20k training setup (37.9$\rightarrow$39.6). In the 1.8M training setup, we obtain 2.1 mAP improvement over the baseline.



To show that our method is generally useful, even in settings where large-scale pretrained model is not available, we conduct experiments without initializing from pretrained checkpoint but otherwise use the same hyperparameters. 
In the non-pretrained setups, we also observe significant improvement on both AS-20k (2.1 mAP improvement) and AS-2M (2.6 mAP improvement), as shown in Table~\ref{tab:performance} (bottom section). 


For the semi-supervised setup, we compare against our 20k supervised method. Without pretraining, we see a 0.6 mAP improvement and a 0.5 mAP improvement with pretraining. 
We conclude that 
consistency regularization can help in an semi-supervise setup using unlabeled data. 

Compared with relevant models listed in Table~\ref{tab:performance} (top section), it is interesting to note that without pretraining, our model does not outperform 
 PANNs~\cite{kong2020large}, probably due to architectural differences. With pretrained checkpoint, our model achieves competitive performance against those of~\cite{gong2021ast} and~\cite{alex2025sslam} which uses similar architectures but different pretraining objectives. 

\section{Conclusion \& Future Work}
In this work, we introduce consistency regularization (CR) into audio event recognition through extensive experiments with AS-20k and AS-2M. Additionally, we apply consistency regularization to a semi-supervised setting. Our findings concluded that 1) consistency regularization in supervised and semi-supervised settings demonstrates consistent improvements across training set sizes, with or without pretraining, 2) multiple number of augmentations with CR increases performance on small datasets. We expect our findings to be generalizable to other sequence modeling tasks. 

\textbf{Future Works:}
With multi-modal models gaining more research focus, Large Audio Language Models (LALM) have room for improvement. We believe that our learned audio representations may lead to further improvements in audio understanding capabilities in LALMs such as LTU\cite{gong2023listen}, SALMONN\cite{tang2024salmonn}, QWEN-AUDIO\cite{chu2023qwen}, and GAMA\cite{ghosh2024gama}
.

\vfill\pagebreak

\bibliographystyle{IEEEbib}
\bibliography{references}


\end{document}